\documentclass[aps,pre,reprint,superscriptaddress,doi=false,isbn=false,url=false,title=true,longbibliography,noeprint]{revtex4-2}
\usepackage{hyperref}
\hypersetup{colorlinks=false, citecolor=blue, urlcolor=blue, linkcolor=blue}
\usepackage{amsmath}
\usepackage{amssymb}
\usepackage{graphicx}
\usepackage{siunitx}
\usepackage[ruled,vlined]{algorithm2e}
\usepackage{todonotes}
\usepackage{cleveref}

\begin{document}
\title{BU-MBAR: A hybrid solution strategy for the MBAR equations}

\author{Fabio M\"uller}
\email{fabio.mueller@neclab.eu}
\author{Francesco Alesiani}
\email{francesco.alesiani@neclab.eu}
\author{Henrik Christiansen}
\email{henrik.christiansen@neclab.eu}
\affiliation{NEC Laboratories Europe GmbH, Kurfürsten-Anlage 36, 69115 Heidelberg, Germany}

\date{\today}

\begin{abstract}
  The multi-state Bennett acceptance ratio (MBAR) equations combine the data collected under different thermodynamic conditions in a statistically optimal way.
  Due to their practical importance, several solution strategies have been devised to optimize the convergence of the resulting set of coupled equations.
  However, even with graphics processing unit (GPU) acceleration, the convergence of these methods can still be either unstable or slow. 
  We propose a new approach where the equations are interpreted as the limit of infinitesimal bin width of the respective weighted histogram analysis method (WHAM).
  In the proposed binned-to-unbinned MBAR (BU-MBAR) method, the bin width is adapted dynamically to ensure a stable and efficient convergence to the asymptotic solution.
  \end{abstract}

\maketitle
\section{Introduction}
\label{sec:introduction}
The multi-state Bennett acceptance ratio (MBAR)~\cite{Bennett1976,Shirts2008} or equivalently the unbinned weighted histogram analysis method (UWHAM)~\cite{Tan2012} represent statistically 
optimal strategies for the weighted analysis of simulation data from multiple trajectories at different thermodynamic conditions.
The solution of the MBAR equations, hence, constitutes an important part of the tool-chain for the post-processing of enhanced sampling data~\cite{mey2020best,Delemotte_2022}. 
The resulting free-energy estimates result from the combination of the statistics from different thermodynamic settings; this allows to reweight to other thermodynamic points, even if no data have been collected explicitly for those conditions.
With growing size of the generated data sets, the solution of the resulting MBAR equation can become increasingly challenging.
Due to the high practical relevance, significant effort has been spent in finding optimized solution strategies.

The MBAR equations are usually presented in a fixed-point form, defining a self-consistent iteration scheme which provides a secure, but potentially slow convergence towards the optimal solution.
Alternatively, the equations can be reinterpreted so that more advanced optimization strategies can be devised.
By employing derivatives, quadratic convergence can be achieved, at least, once the solution is sufficiently close.

One of the most established implementations is provided by the PYMBAR package~\cite{Shirts2008}, which offers a variety of different solution strategies based on combinations of self-consistent iterations and derivative-based approaches.
Ref.~\cite{Ding2019} developed an implementation in PyTorch~\cite{paszke2017automatic}, providing graphics processing unit (GPU) acceleration.
It was shown that a speedup of $\approx 150$ compared to PYMBAR can be achieved for large-scale problems when running on the GPU.
However, by now, PYMBAR has been optimized for GPU execution using JAX~\cite{jax2018github}, achieving similar performance when running on GPUs.

In a later work, a ``divide and conquer'' approach was proposed for the solution of the MBAR equations~\cite{Jia2021}.
There, no global solution of the equations is sought, but rather the equations are solved for subsets of replicas whose mutual overlap surpasses an arbitrarily fixed threshold value.
Despite reducing the computational complexity from quadratic to linear in the number of thermodynamic states, it is claimed that the global solution of the full MBAR equations is well approximated for a sufficiently small threshold.
A more recent solver adopts a stochastic approach akin to the training of deep neural networks~\cite{Galama2023}, where iterations are performed with randomly chosen batches of the full data set.
However, this approach is not hyperparameter-free and requires a protocol that specifies the evolution of the learning rate and the batch size during the solution process.

Several modifications of the MBAR equations have been proposed to target specific problems related to the reweighting procedure.
One potential problem is given by the presence of multiple free-energy minima separated by free-energy barriers.
This can lead to very long transition times from one basin to another, such that the data within the corresponding timeseries have very long-lived relaxation modes.
To address this problem, a stratified version of the MBAR equations was developed, where the transition rates between the different basins enter the equations~\cite{Zhang2017,Zhang2019}, correcting for possible biases.
Another challenge is the proper assessment of confidence intervals for the free-energy estimates, which, for example, is tackled by the BayesMBAR approach~\cite{Ding2024}.
It is shown that for systems with few degrees of freedom, the resulting error estimates are more reliable than conventional error estimates as, e.g., obtained by bootstrapping~\cite{Efron1982}.

The numerous approaches targeting the improvement of the MBAR formulation and the solution of the resulting equations illustrate the significance of a robust and fast solution strategy which can deal with large-scale MBAR problems in reasonable timescales.
In this work, we propose a strategy for the solution of the MBAR equations by regarding them as the limit of the corresponding weighted histogram analysis method~\cite{Ferrenberg1989, kumar1992weighted} (WHAM) equations, where the limit of infinitesimal bin width is taken dynamically.
The resulting {\it binned-to-unbinned MBAR} (BU-MBAR) solver is both fast and robust, outperforming a direct solution of the unbinned equation in terms of wall-clock time.
An open-source GPU implementation is available at \url{https://github.com/nec-research/BU-MBAR}.

In the following, we first present the algorithmic details of the BU-MBAR method in \cref{sec:implementation}.
Then, we apply the method to a parallel tempering simulation of alanine dipeptide in explicit solvent and compare the results to PYMBAR in \cref{sec:dialanine}.
Finally, we investigate the scaling of the execution time with respect to the number of replicas for synthetic data in \cref{sec:synthetic} before concluding with a discussion and outlook in \cref{sec:discussion}.

\section{Algorithm}
\label{sec:implementation}
We start by recalling the WHAM equations for the combination of data from $m$ different timeseries from simulations at the corresponding inverse temperatures $\beta_i = 1/kT_i$ with $i=1,\ldots,m$. For each temperature, we sample the energies $E_{il}, l=1,\dots, N_i$, with $N_i$ being the number of measurements for $\beta_i$.
Ignoring any influence of the autocorrelations on the effective statistics~\cite{Janke2013} the equations for the free energy estimates $f)i^{k}$ at iteration step $k$ read
\begin{equation}
\label{eq-iterative}
  f^{k+1}_i = -\log\left[ \sum_{p=1}^{n} \frac{H^{(n)}_p e^{- E_p \beta_i}}{\sum_{j=1}^m N_j e^{f^{k}_j - E_p \beta_j}} \right],
\end{equation}
defining a self-consistent iteration (SCI) scheme where the right hand side is calculated with the values $f_i^{k}$ in iteration step $k$ which result in the new values $f^{k+1}_i$ in iteration step $k+1$ on the left hand side.
$H^{(n)}_p=\text{Hist}^{(n)}(E_p)$ denotes the number of entries for the bin with bin center $E_p$, where $n$ is the number of nonempty bins of the histogram. 
Since bins where $H^{(n)}_p=0$ are ignored in the evaluation of (\ref{eq-iterative}), the index $p$ only includes nonempty bins, i.e., $H_p^{(n)} \ge 1, \forall p \in \{1, \dots, n\}$.
The frequency of empty bins depends on the chosen bin width; the target number of nonempty bins can only be achieved with a correspondingly finer bin width.
A simple approach is to define the width $w_p=w_p(n)$ of each bin such that it is the smallest uniform width that generates $n$ nonempty bins.
In this case, the number of nonempty bins is a function of the uniform width, i.e., $n=g(w)$, and we select $w$ such that $w^*(n) = \max_{w: n \le g(w)} w$.
In the limit of infinitely many bins each (unique) measured $E_p$ ends up in its own bin, 
recovering the UWHAM equations where the summation is performed over the individual measurements.
\par
With a growing number of replicas, when relying solely on the self-consistent iteration scheme, increasingly many steps are necessary to achieve the desired accuracy of the solution.
A remedy to this problem, which works well sufficiently close to the optimal solution, is the use of derivative-based optimization strategies that promise quadratic convergence (i.e., doubling of the correct digits in each iteration step)~\cite{Shirts2008}, see Ref.~\cite{Press2007} for different approaches. 
Since the final convergence becomes very fast using these advanced techniques, most of the effort is spent in exploring the global loss landscape of the equations to find the basin of attraction where quadratic convergence can be achieved.
\par
Our contribution is a new approach to reach an approximation to the global minimum in a drastically reduced time.
Subsequently, one can switch over to derivative based or rely on other established solver such as PYMBAR~\cite{Shirts2008}.
The idea starts from the observation that the UWHAM equation can be seen as the limit of infinitesimally small bins of Eq.~(\ref{eq-iterative}).
For each number of nonempty bins $n$, which we assume we can achieve using a constant histogram bin width $w(n)$, $f_i(n)$ converge to a different $f^*_i(n)$.
Only for the limit of the infinitesimal bin width, i.e., each measurement ends up in a separate nonempty bin the iteration converges to the correct fixed point $f_i^*(N)$.
To measure the progress per iteration step, we introduce the relative step width in dependence on the number of bins $n$
\begin{equation}
\label{eq:stepwidth}
  s(f_i, n) = \mathrm{max}\left[\left|\frac{f_i^\mathrm{new}(n) - f_i^\mathrm{old}(n)}{f_i^\mathrm{old}(n)}\right|\right],
\end{equation}
where $\left| \ldots \right|$ denotes the element-wise absolute, $f_i^\mathrm{old}(n)$ is the current estimate for $f_i(n)$, and $f_i^\mathrm{new}(n)$ is the new estimate.
Since the outer sum in (\ref{eq-iterative}) runs over all bins, it immediately follows that the cost of one self-consistent iteration step scales linearly with $n$ so that it is computationally much cheaper to perform single iterations with $n \ll N$.
However, since for $n<N$ the iteration converges to a wrong fixed point, it is not useful to iterate the equation up to $f^*(n)$.
A suitable histogram resolution is imposed by periodically considering a larger bin number $n'$ and comparing $s(f_i, n)$ and $s(f_i, n')$.
Heuristically, the number of bins is increased to $n'$ when
\begin{equation}
  \label{eq:heuristic}
  s(f_i, n)/n < 2s(f_i, n')/n',
\end{equation}
is fulfilled.
The idea behind this heuristics is that the step width normalized by the computational cost (which is proportional to $n$ itself) should be maximal, with a small bias (the factor of two on the r.h.s.) towards larger $n$, for which we know that the systematic error is smaller.
While this self-consistent criterion forms the basic philosophy of the new approach, which is in principle sufficient for an efficient implementation, we also provide a more detailed algorithmic description as pseudo-code in Algorithm~\ref{alg:BU-MBAR}.

\newcommand{\inc}[0]{\mathrm{inc}}
\begin{algorithm}[t]
\DontPrintSemicolon
\KwIn{$\beta_i$ (inverse temperature), $N_i$ (number of measurements per temperature), $i=1,\dots,m$, with $m$ (number of temperatures)}
\textbf{Settings:} $n_0$ $\epsilon$, $\inc$, 
$N_\text{intermediate}$ (number of intermediate SCI steps), 
$\text{SCI}(\mathbf f, n)$ finds the self-consistent solution with $n$ nonempty bins using \cref{eq-iterative}
\;
\KwOut{Converged solution vector $\mathbf{f}=(f_1,\dots,f_m)$}
Initialize $\mathbf{f} \gets \mathbf{0}$\;
$n \gets n_0$\;

\While{True}{
    $\mathbf{f}_{new} \gets \text{SCI}( \mathbf f, n)$\;
    $s \gets \max(|\mathbf{f}_{new}/\mathbf{f} - 1|)$\;
    $\mathbf{f} \gets \mathbf{f}_{new}$\;
    \If{$s < \epsilon$}{ \textbf{break} }
}

$n \gets n \times \inc{}$\;

\While{True}{
  \For{$i=1$ \KwTo 
  $N_\text{intermediate}$
  }{
    $\mathbf{f} \gets \text{SCI}( \mathbf f, n)$\;
  }
  $\mathbf{f}_{new} \gets \text{SCI}( \mathbf f, n)$\;
  $s_n \gets \max(|\mathbf{f}_{new}/\mathbf{f} - 1|)$\;
  $\mathbf{f}_{next} \gets \text{SCI}( \mathbf f, n \times \inc{})$\;
  $s_{next} \gets \max(|\mathbf{f}_{next}/\mathbf{f} - 1|)$\;
  
  \If{$2 \cdot s_{next} > \inc{} \cdot s_n$}{
    $n \gets n \times \inc{}$\;
    \If{$2 \cdot \inc{} \cdot n > \sum N_i$}{ \textbf{break} }
  }   
}

\caption{Binned-to-Unbinned MBAR (BU-MBAR): an adaptive convergence algorithm.}
\label{alg:BU-MBAR}
\end{algorithm}

The actual algorithmic implementation requires a set of hyper parameters whose values were kept fixed for all the benchmarks reported in this work:
the initial number of bins $\mathrm{n_0}=100$, the step size up to which iteration with $n_0$ bins are performed $\epsilon=10^{-3}$, the factor by which target number of bins is increased $\inc{}=10$, and the number of self-consistent iterations between subsequent checks of self-consistent criterion~(\ref{eq:heuristic}) for increasing the number of bins $N_\text{intermediate}=20$.
We observe that, the algorithm's performance is robust even with respect to significant changes of these settings and did not require any adjustment for any of the cases considered here.

\section{Alanine dipeptide}
\label{sec:dialanine}
One typical application where the solution of the MBAR equation is needed is the combination of statistics gathered from parallel tempering (PT) simulations \cite{Swendsen1986, Geyer1992, Hansmann1997, Sugita1999}.
PT, also referred to as temperature replica exchange, is one of the most versatile generalized ensemble techniques where different replicas are simulated at different temperatures.
In addition to evolving individual replicas through steps of molecular dynamics (or alternatively through Monte Carlo moves), pairs of thermodynamic states $i$ and $j$ are allowed to exchange their configurations with probability
\begin{equation}
  \label{eq:exchange-probability}
  p = \mathrm{min}\left( 1, e^{(E_i - E_j)\left(\beta_i - \beta_j \right)} \right),  
\end{equation}
where $E_i$ and $E_j$ are the potential energies of the configurations of the thermodynamic states $i$ and $j$, respectively, and $\beta_i$ and $\beta_j$ the corresponding inverse temperatures.
In case of MD, instead of being explicitly considered, kinetic energies are only re-scaled to the correct value.
The intention behind PT is to allow the system to surmount free energy barriers at higher temperatures while exploring relevant free energy valleys at lower temperatures.
Each time a configuration visits the highest temperature, it can be expected to completely decorrelate.
\par
Here, we perform simulations of alanine dipeptide described by the amber14SB force field, solvated in 863 rigid TIP3P water molecules using OpenMM 8.4~\cite{Eastman2024}.
The system is contained in a cubic periodic simulation box whose size, after an initial relaxation at $T=\qty{298}{\kelvin}$ and 1\ atm, is kept fixed at $(\qty{2.97}{\nm})^3$.
The ensemble is composed of 50 replicas with exponentially spaced temperatures between $\qty{298}{\kelvin}$ and $\qty{600}{\kelvin}$~\footnote{This corresponds to the standard temperature schedule of OpenMM where the temperatures are chosen as $T_\mathrm{min} \times \exp(T_i)$, with $T_i \in [0, \log (T_\mathrm{max}/T_\mathrm{min})]$ being $N$ equidistant temperature points on the closed interval.}.
The temperature is controlled by a Langevin thermostat with collision rate $\qty{1}{\per\ps}$ and stepsize $\qty{2}{\fs}$.
After an initial equilibration period of $\qty{2}{\ns}$, each trajectory is simulated for $\qty{0.2}{\us}$ with measurements and all-to-all replica exchange attempts performed every $\qty{1}{\ps}$, resulting in a total of 200\,000 measurements per replica.
Long-range interactions are handled with particle-mesh Ewald, and for Lennard-Jones interactions a cutoff of $\qty{0.9}{\nm}$ and switching distance $\qty{0.85}{\nm}$ was used.
The hydrogen bonds are constrained using CCMA~\cite{Eastman2010} and the motion of the center-of-mass is removed.
\par
\begin{figure}
  \centering
  \includegraphics[width=\columnwidth]{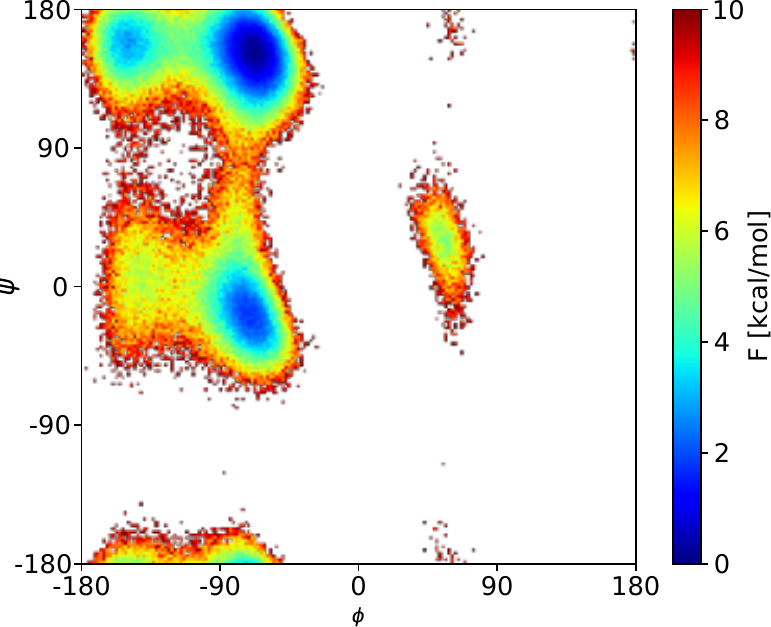}
  \caption{MBAR estimator of the free-energy profile of the alanine dipeptide projected onto the torsion angles $\phi$ and $\psi$ at $\qty{298}{\kelvin}$.
    The data was generated in a parallel tempering simulation with 50 replicas in the temperature range $\qtyrange{298}{600}{\kelvin}$.
    }
  \label{fig:free-energy}
\end{figure}
An established way to visualize the ensemble are free energy plots, where the logarithm of the probability density is plotted after projecting the configuration onto some appropriate collective variables capturing the slow modes of the system, which in this case are given by the pair of backbone dihedral angles $\phi$ and $\psi$.
Using the self-consistent solutions $f_m$ of the MBAR equations, the statistical weight of each individual measurement at the inverse temperature $\beta_0$ can be calculated as
\begin{equation}
  \label{eq:individual-probability}
  p_i = \frac{1}{C} \frac{e^{-\beta_0 E_i}}{\sum_{m=1}^M N_m e^{f_m-\beta_m E_i}},
\end{equation}
with the normalization constant,
\begin{equation}
  \label{eq:normalization}
  C = \sum_{i=1}^N p_i,
\end{equation}
where $N = \sum_{i=1}^m N_i$ is the total number of measurements from all different temperatures. 
The corresponding plot is shown in Fig.~\ref{fig:free-energy}, where a binning width of $\qty{2}{\degree}$ was chosen for both dihedral angles.
\par
A prerequisite for the evaluation of (\ref{eq:individual-probability}) and (\ref{eq:normalization}) are the solutions to the MBAR equations, i.e., the relative free energies $f_i$.
\begin{figure}
  \centering
  \includegraphics[width=1.05\columnwidth]{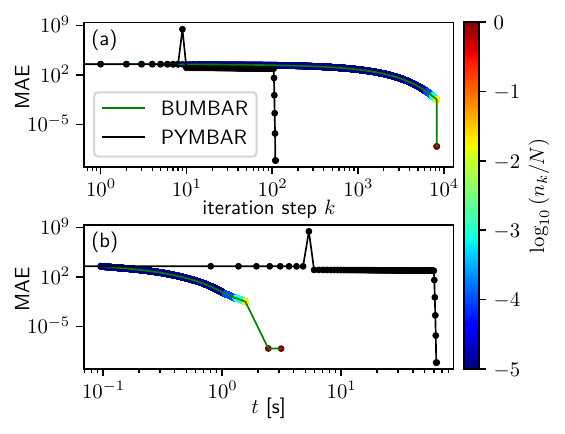}
  \caption{Evolution of the mean-absolute error (MAE) of the relative free energies for the full dataset generated from the PT simulations for BU-MBAR and PYMBAR.
    Evolution shown in dependence of (a) iteration steps (b) wall-clock time.
    Color encodes the number of histogram bins used by BU-MBAR at iteration step $k$.}
  \label{fig:progression}
\end{figure}
Using BU-MBAR, the solution of the equations takes less than $\qty{4}{\s}$ compared to more than $\qty{60}{\s}$ using PYMBAR with the ``robust'' setting and starting with $\mathbf{f}_i=\mathbf{0}$ as initial guess.
The direct comparison of the progress of the solution is analyzed in Fig.~\ref{fig:progression}, where the quality of the solution is shown in terms of the mean-absolute error MAE with respect to the final solution $f^*_{i,\mathrm{PYMBAR}}$ found by PYMBAR,
\begin{equation}
  \label{eq:mae}
  \mathrm{MAE} = \frac{1}{M} \sum_{i=1}^{M} \left| f^*_{i,\mathrm{PYMBAR}} - f_i \right|.
\end{equation}
In Fig.~\ref{fig:progression}~(a), the progression towards the solution is presented in dependence on the iteration step.
PYMBAR needs roughly 100 iteration steps to find the final solution, while BU-MBAR needs roughly $10^4$ iterations.
The color code represents the number of nonempty histogram bins in the iteration step $k$, showing that most of the iteration steps are performed with very few histogram bins.
The wall-clock time is analyzed in panel (b) with both methods executed on an NVIDIA A40 GPU and discarding the initial compilation time for PYMBAR.
Here, it becomes apparent that, for BU-MBAR, the last two iterations using a quasi-Newton scheme with unbinned data take up roughly as much time as all previous thousands of self-consistent iterations with binning.
In the initial phase, the additional systematic error from the coarser histogram is negligible, such that the algorithm keeps on performing more iterations with unchanged binning.
Only once the systematic error of the binning becomes comparable to the step size towards the true solution a finer binning will be chosen.
Hence, in the case of more difficult problem instances, the algorithm will mainly need more initial iteration steps with coarse binning which barely contribute to the total execution time.
For PYMBAR, instead, all iterations take a similar amount of time, rendering the total wall-clock time proportional to the number of iterations needed to reach the final solution.
This implies that PYMBAR's execution time may strongly depend on the hardness of the underlying optimization problem, while this dependency can be expected to be less pronounced for BU-MBAR.
The dependence of the execution time of the algorithm on the hardness of the problem is investigated in the next section using synthetic data.

\section{Synthetic Data}
\label{sec:synthetic}
\begin{figure}
  \centering
  \includegraphics{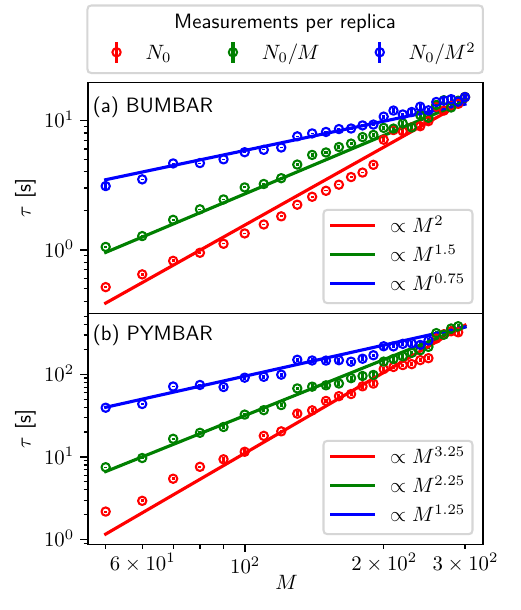}
  \caption{Wall-clock time scaling for (a) BU-MBAR in comparison to (b) PYMBAR in dependence of the number of replicas $M$.
    For both methods exactly the same energy histograms are generated and both methods are run on the same GPU.
    The three different colors correspond to three different scaling of the number of measurements per replica.
    The lines are intended as a guide to the eye.
    }
  \label{fig:scaling-synthetic}
\end{figure}
One problem with PT simulations in explicit solvent is the necessary growth of the simulation box with increasing size of the biomolecule studied.
The growth of the simulation box implies an even faster growth of the amount of solvent and hence the number of degrees of freedom of the system, causing a narrowing of the energy histograms. 
To cover the same temperature range, the reduced width of the histograms needs to be compensated for by a larger number of intermediate temperature points to maintain a reasonable overlap between neighboring histograms, necessary for sufficient exchange probabilities between adjacent replicas.
Here, we emulate this situation by randomly drawing potential energies from normal distributions of unit variance for which we impose a constant overlap between neighboring histograms of 80\%.
This allows us to easily generate an arbitrary number of temperature points and samples to obtain a more detailed impression of the algorithmic improvement of the method.
\par
Following this approach, we investigate three different scenarios, always keeping the number of samples per replica $N_i=N_0=\mathrm{const.}$ equal among all different replicas:
i) The number of uncorrelated samples per replica $N_0(M) = \mathrm{const}$ does not depend on the number of replicas $M$, i.e., the total number of samples goes as $N \equiv \sum_{i=1}^M N_0 \sim M$ and also the involved simulation effort is proportional to $M$~\footnote{When neglecting any algorithmic overhead and the interplay between $M$ and the integrated autocorrelation time, theoretically, for this scenario the wall-clock time of the simulation can be kept constant if each replica is simulated on a separate GPU.}. A single self-consistent iteration step using the unbinned MBAR equation needs $\sim M^2$ operations since each of the $N$ measurements must be evaluated at all $M$ different temperature points.
ii) The number of samples per replica is chosen to be $N_0/M$, leading to $N(M) = \mathrm{const}$ and $\sim M$ cost per iteration. This scenario corresponds to constant simulation time if all simulations are run serially on a single device.
iii) The number of samples per replica is chosen to be $N_0/M^2$, leading to $N \sim 1/M$ and a constant cost per self-consistent iteration step.
Specifically, we set $N_0=10^4$ for the red curve, $N_0=3\times 10^6$ for the green curve, and $N_0=9\times 10^8$ for the blue curve, resulting in $10^4$ samples for $M=300$ for all three cases.
This choice of the problem size is mainly driven by the memory requirements, which limit the maximum problem size in a conventional vectorized implementation where the full energy matrix at all different thermodynamic points is constructed.
The resulting timings are presented in Fig.~\ref{fig:scaling-synthetic} (a) for BU-MBAR and (b) for PYMBAR.
Both solvers were executed on the same NVIDIA A40 GPU using the same data.
Each setting was averaged over 12 different runs with different random number generator seeds.
In a direct comparison between the two methods, BU-MBAR has reduced runtimes for all settings.
As a guide to the eye, we draw power laws that roughly follow the observed computational complexity.
For all three different settings, the computational complexity for PYMBAR matches $M^{1.25}$ multiplied by the computational complexity of the cost for one single self-consistent iteration step.
For BU-MBAR, the computational complexity cannot be understood within such a simple scheme.
Nonetheless, the algorithmic scaling is reduced for all three considered cases, implying that even larger speedups can be expected for larger problem sizes.

\section{Discussion and Conclusion}
\label{sec:discussion}

We have presented a new method for the solution of MBAR equations.
By regarding the MBAR equation as the limit of infinitely small histogram width of the WHAM equations, the iterations are performed starting from a small number of bins and going to the limit of infinitesimal bin width.
The histogram resolution is self-consistently tuned to maximize the convergence rate per fixed computation effort.
The approach retains the numerical stability of the self-consistent solution strategy, reliably reaching an initial solution, which is in the regime where the equations behave nearly linearly, and thus a quadratic convergence with the gradient-based method can be achieved.
Hence, the number of iteration steps at full resolution and using the gradient based methods becomes nearly independent of the numerical difficulty of the original problem.
For the scenario considered here of parallel tempering ensembles with a varying number of replicas at different temperatures and varying statistics, we see in each case a reduction of the scaling of the computational cost with the number of replicas. 
\par
In the solution of the MBAR equations the bottleneck is mainly constituted by the very high cost of the single function evaluation.
By considering the binned form of the equations (WHAM) a systematic error is introduced, i.e., the fixed point of the equations is shifted.
By self-consistently balancing this systematic error with the rate of convergence and the computational cost, we could find an efficient solution strategy which is very robust with respect to the parameter of the algorithm and does not need any manual adjustment for different problem cases.
Despite the speedup, the correctness of the final solution is guaranteed by reaching the limit of infinitesimal bin width.
A future perspective could be the generalization of the binned-to-unbinned to differently defined generalized ensembles.
It would also be interesting to investigate whether a similar strategy can render schedules for stochastic optimization approaches~\cite{Galama2023} more robust, by seeing the batch size in correspondence to the number of histogram bins.

\section*{Data Availability}
The source code of the BU-MBAR solver, including the alanine-dipeptide simulation and analysis scripts and the synthetic energy-overlap benchmark used in this work, is available at \url{https://github.com/nec-research/BU-MBAR}.

\bibliography{./bibliography.bib}

\end{document}